# Planetary Nebulae Shaped By Common Envelope Evolution

Adam Frank [1],*, Zhuo Chen[1], Thomas Reichardt[2], Orsola De Marco[2], Eric Blackman[1] and Jason Nordhaus[3]

[1]*Department of Physics and Astronomy, University of Rochester, Rochester, NY 14627, USA;*
[2]*Department of Physics & Astronomy, Macquarie University, Sydney, NSW 2109, Austrialia*
[3]*National Technical Institute for the Deaf, Rochester Institute of Technology, NY 14623, USA*
* Correspondence: afrank@pas.rochester.edu

**Abstract:** The morphologies of planetary nebula have long been believed to be due to wind shaping processes in which a "fast wind" from the central star impacts a previously ejected envelope. Asymmetries assumed to exist in the "slow wind" envelope lead to inertial confinement shaping the resulting interacting winds flow. We present new results demonstrating the effectiveness of Common Envelope Evolution (CEE) at producing aspherical envelopes which, when impinged upon by a spherical fast stellar wind, produce highly bipolar, jet-like outflows. We have run two simple cases using the output of a single PHANTOM SPH CEE simulation. Our work uses the Adaptive Mesh Refinement code AstroBEAR to track the interaction of the fast wind and CEE ejecta allowing us to follow the morphological evolution of the outflow lobes at high resolution in 3-D. Our two models bracket low and high momentum output fast winds. We find the interaction leads to highly collimated bipolar outflows. In addition, the bipolar morphology depends on the fast wind momentum injection rate. With this dependence comes the initiation of significant symmetry breaking between the top and bottom bipolar lobes. Our simulations, though simplified, confirm the long-standing belief that CEE can plan a major role in PPN and PN shaping.

## 1. Introduction

Planetary nebulae ("PN") are formed via gas ejected from highly evolved stars starting within a few thousand years before their final state as white dwarfs. The interaction between a fast wind ejected during the post-AGB evolutionary state of the star and the previously ejected AGB slow material drives both Pre-PN ("PPN") and mature PN (Balick & Frank 2002). Given the bipolar nature of many PPN and PN, a "Generalized Interacting Stellar Wind" model has long been thought to be a dominant mechanism for understanding their shaping. The Interacting Stellar Winds model began with the seminal work of Kwok et al (1978) who proposed that fast, line driven winds from the hot Central Star of a PN (CSPN) expanding at velocities of order $v_f \sim 1000$ km/s would drive shocks into the heavy and slowly expanding AGB wind ejected earlier ($v_f \sim 10$ km/s). The bright rims of PN were then the ionized shells of swept up AGB material. The generalization of this model to explain aspherical PN began with suggestions by Balick (1987) and Icke (1988) that slow AGB winds could include a pole-to-equator density contrast. Once a spherical fast wind begins impinging on this aspherical slow wind, "inertial confinement" leads the resulting shocks to take on elliptical or bipolar configurations. In this Generalized Interacting Stellar Wind model (GISW), the run from mildly elliptical PN to strongly bipolar "butterfly" shaped nebula depends on the pole-to-equator density contrast in the AGB slow wind (e = $\rho_P / \rho_e$) and its aspherical morphology.

Analytic solutions (Icke 988) and later simulations (Icke et al 1992, Frank et al 1993, Frank and Mellema 1994) verified the ability of the GISW model to capture observed PN morphologies. With time, other shaping processes were considered including the role of magnetic fields (Garcia-Segura et al 1999, 2005), clumps (Steffen et al 2004) and the possibility that the fast wind was already collimated in the form of jets (Lee & Sahai 2004). In addition, new observations indicated that much of the shaping process for PN may actually occur before the central star heats up enough to produce ionizing radiation (the post-AGB/PPN phases Bujarrabal et al 2001). While this recognition, along with the other mechanisms such as MHD, have contributed greatly to the understanding of PN shaping, the GISW model often remains a background aspect of their scenarios. For example, MHD models may still require a strong pole to equator density contrast to produce certain kinds of wasp-waisted nebula. Thus in spite of the evolution of PN shaping studies since Kwok, Balick and Icke's initial ground-breaking studies, the existence of a pre-existing toriodal density distribution still figures centrally in our understanding of the PN shaping process.

Of course, the assumption of a toriodal circumstellar slow wind has always begged the question of its formation. While single star models for the production of a pole-to-equator density contrast have been proposed (Bjorkman et al 1993), most of these rely on degrees of rotation that have proven difficult to obtain for slow rotating AGB stars. Thus binary stars have been seen as essential to modeling PN shaping for some time. Early work of Livio and Soker (Livio 1994, Soker 1989) provided key insights into how binary interactions could produce aspherical circumstellar environments. More recent 3-D simulation work has shown how even detached binary interactions can produce fertile ground for PPN shaping (Chen et al 2016).

Throughout the extensive discussion of the role that binary stars can play in creating PN, Common Envelope Evolution (CEE) has long been seen a primary means for generating AGB environments with a high pole to equator density contrast (Livio & Soker 1988 ). Common Envelope Evolution occurs when a more compact companion plunges into the envelope of an RGB or AGB star (Paczynsk, 1976). The release of gravitational energy during the rapid orbital decay is expected to unbind some, or all, of the AGB envelope creating an expanding toroidal flow that can then serve as the aspherical slow wind for GISW models. The ability to calculate CEE out to the point where the expanding envelope could serve as input for GISW models has, in the past, been hampered by the complexity of CEE 3-D flows. Over the last decade, however, a number of new simulation platforms have been developed for CEE including smooth particle methods (SPH **Passy et al 2012**), AMR fixed mesh methods (Ricker et al 2012, Chamony et al 2018) as well as moving mesh methods (Olhmann 2015).

In this contribution we take on the problem of PPN evolution from CEE systems by combining a SPH CEE simulation with an AMR grid-based wind interaction simulation. Our goal in these initial studies is map out the basic flow patterns emerging from GISW interactions with CEE initial conditions using fully 3-D, high-resolution simulations. We note that 2-D fixed grid studies of this problem have recently been completed by Garcia-Segura et al 2016 using a different set of CE inputs.

**2. Model and Methods**

Our simulations begin with the output of CEE models calculated using the SPH code PHANTOM (Price et al. 2018). The model tracks a binary system with a $M_1 = 0.88\ M_\odot$ primary and $M_2 = 0.6\ M_\odot$ companion. The envelope of the primary holds $M_{1,e} = 0.48\ M_\odot$ of mass. Thus there is approximately 1.1 $M_\odot$ in the two stellar cores that will eventually form a tight binary or merged object. From the simulations the final separation of the binary has stabilized at approximately $R = 20\ R_\odot$ (for more details on the simulation see Iaconi et al. 2017 and De Marco et al. 2017). The results in terms of 3-D flow variable distributions then become the initial conditions for our AstroBEAR AMR simulation (for details of the AstroBEAR AMR multi-physics code see Carroll-Nellenback et al 2013). The numerical results from the SPH simulation are the sampled at points that fit into an Eulerian mesh with 7 levels of AMR. The base grid of the mesh has a physical dimension of $(1000\ R_o)^3$ and the 7th level grid has a physical dimension of $(7.8\ R_o)^3$.   The simulation domain is cropped to $(16000\ R_o)^2 \times (128000\ R_o)$. The origin of the simulation is set at the center of the domain.

After reading in the original SPH data and mapping it to the Eulerian mesh, a point particle with mass 1 $M_o$ is placed at the grid origin.   The point particle's gravity is imposed on the gas however the self-gravity of the gas is not considered. A spherical inner "wind" boundary is created around the point particle with r = 46.9R_sun (spanning 6 of finest mesh cells). The fast wind conditions are injected into the grid through this boundary. Note that the binary orbit at the end of the SPH CE simulation fits within this wind boundary region.    Thus we do not resolve the binary's evolution any further or consider its effect on the fast wind that is considered a spherical outflow in our models.

| Table 1. | Case A: High Momentum Flux | Case B: Low Momentum Flux |
|---|---|---|
| $\rho_{FW}$ (g cm$^{-3}$) | $1.0 \times 10^{-11}$ | $5.0 \times 10^{-13}$ |
| $V_{FW}$ (km s$^{-1}$) | 300 | 300 |
| Quiescent Phase $\Delta t$ (days) | 500 | 6000 |
| Mass Loss Rate ($M_o$ y$^{-1}$) | $6.4 \times 10^{-4}$ | $3.2 \times 10^{-5}$ |
| $T_{FW}$ (°ak) | 30000 | 30000 |

After preparing the AMR grid with the SPH CEE conditions, we allow for a quiescent period of evolution where the CEE ejecta is simply allowed expand for some time before the fast wind is initiated. We do this to allow the expanding CE flow to relax on the grid.   We note that while we do not include self-gravity in the calculation, we do not see significant evolution of the morphology of the CE ejecta.   The most important feature of the quiescent period is that some innermost CE material falls back onto the inner boundary condition (which during this period is set to allow for infall).    The fast wind is then turned on instantaneously.    We run two different models for the fast wind: a low and a high momentum flux case. We list both type of condition in the table 1.

3. Results

In the GISW model two shocks form as the fast wind interacts with the previously deposited slow moving material.    An outer shock, facing radially outward from the central wind source, is created as slow wind material is accelerated and swept up via the driving action of the interior fast wind. In addition an inner shock forms facing back to the wind source via the deceleration of the fast wind. The inner shock converts wind kinetic energy into thermal energy of post-shock fast wind material. If both the fast and slow wind are spherical, the two shocks will be spherical.    If, however, the slow wind is aspherical the inner shock takes on a convex semi-elliptical shape. This distortion of the inner shock is important because radial fast wind streamlines no longer encounter the shock normally but instead strike it an angle. Streamlines passing through a shock at an acute angle are refracted away from the shock normal. In this way an aspherical slow wind in GISW models leads to a "shock focused inertial confinement of the fast wind producing well collimated jets (Icke et al 1992). The stronger the pole to equator contrast in the slow wind the more effective is the shock focusing of fast wind material into bipolar jets (Frank et al 1999).

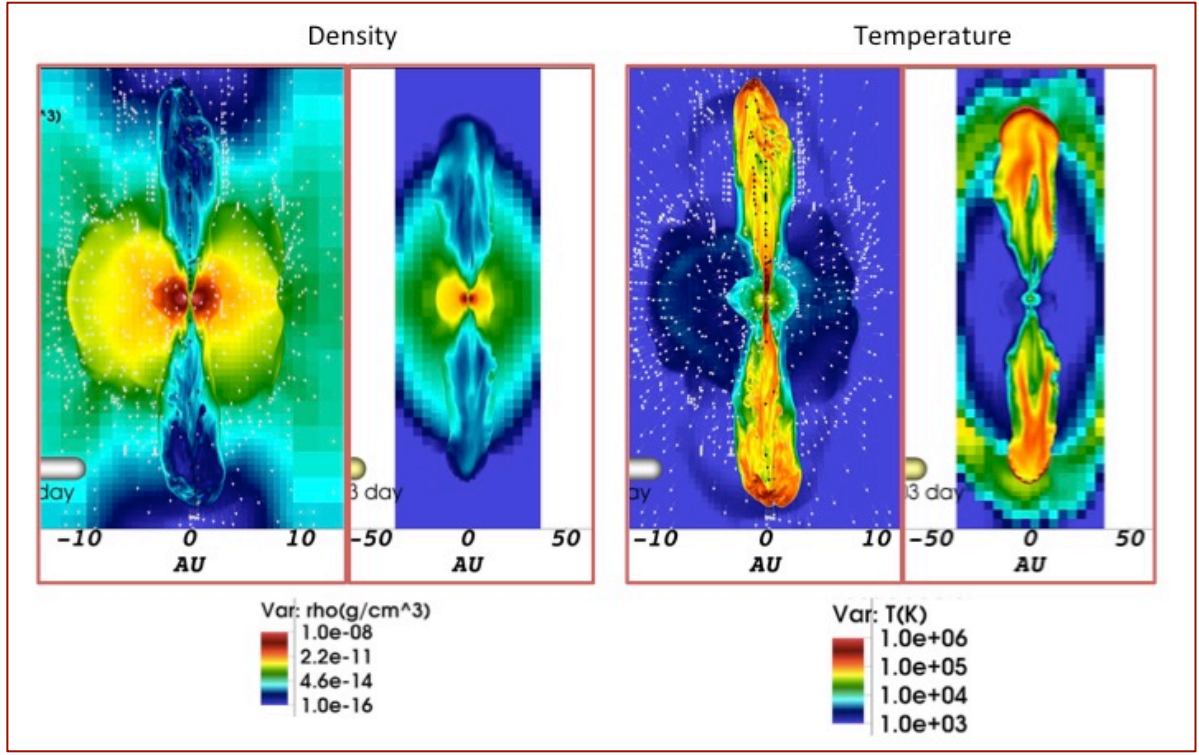

**Figure 1: 2-D slices in the X-Z plane of log density (right) and log temperature (left) from 3-D AMR simulations of a high momentum fast wind being driven into CEE ejecta. We show two different times for the simulations (each left sub-panel is at t = 780 days after initiation of fast wind. Each right sub-panel is at t = 1600 days after initiation of fast wind. Note the change of scale between the two times as the nebular lobes expand.**

*Model A:* In figure 1 we show slices in the x-z plane from our AMR AstroBEAR simulation taken at two different times. The first slice is taken 780 days after the initialization of the fast wind and the second is taken at 1600 days.

We first note that the CEE ejecta began with a high pole to equator density contrast (e > 100) and a "cored apple" morphology. During the quiescent evolution of the CEE ejecta we did not observe this morphology to change significantly. Once the fast wind begins we see that the high densities of the CEE lead to strong inertial confinement in which the momentum injection via the fast wind is unable move ejecta material in the equatorial direction. In the polar regions, however, fast wind material is able to drive through the ejecta leading to two highly collimated shocked wind flows which, together, create an overall bipolar outflow.

Consideration of the flow at t = 780 days shows that the lobes are essentially hollow, bounded externally by shocked, swept up CEE ejecta and filled internally with low density shocked fast wind material. The shocked fast wind leads, as one would expect, to high temperatures (T ~ $10^6$ K) in the lobe interiors.

As the lobes expand, their basic morphology remains relatively steady with a length to radius ratio of order L_lobe/R_lobe  7. What is noteworthy however is the difference between the upper and lower lobes in the two time slices.   Because the evolution of the CE ejecta need not respect top/bottom symmetry relative to the orbital plane, it presents a different environment for fast wind interaction above and below the equatorial regions.   Thus we see the details of density and temperature inhomogeneities in the top and bottom lobes differing significantly.   This natural cause of symmetry breaking in the two lobes holds great promise in explanatory the rich morphologies seen in high-resolution images of PNe.

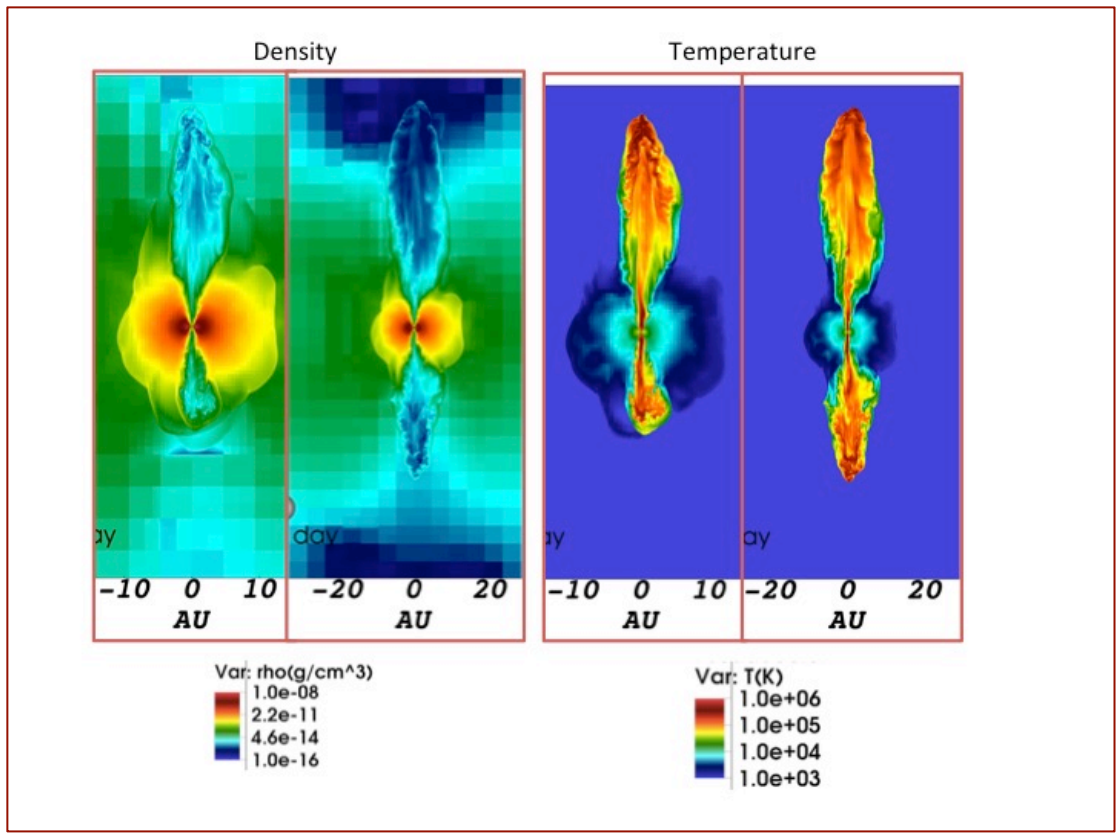

**Figure 2: 2-D slices in the X-Z plane of log density (right) and log temperature (left) from 3-D AMR simulations of a low momentum fast wind being driven into CEE ejecta.   We show two different times for the simulations (each left sub-panel is at t = 6440 days after initiation of fast wind. Each right sub-panel is at t = 6770 days after initiation of fast wind.   Note the change of scale between the two times as the nebular lobes expand.**

*Model B:*   In figure 2 we show slices from a simulation with a fast wind momentum injection rate that is 20 times lower than Model A.   In this model it takes considerable time for the bipolar lobes to work their way out of the aspherical CEE ejecta as is demonstrated by the time at which the two sets of images are taken (t = 6440 and 6770 days respectively).   The two times shown are relatively close, as compared to what was shown in Model A.   This is because before t = 6440 the lobes have yet to breakout from the central regions of the CEE ejecta.   Once the lobes can escape the central regions they begin a rapid expansion in the z-direction (as defined by the binary orbital angular momentum vector).

What is most noteworthy for this model is the strong asymmetry between the top and bottom lobes. While Model A showed differences in details of density/temperature inhomogeneities, Model B provides evidence for a very different evolutionary history between the two lobes. While the top lobe has advanced to almost Z = 60 AU from the central source, the bottom lobe has only made it to Z = 40 AU. Inspection of the evolution of the Model B shows that the difference between the top and bottom lobe lies in sensitivity to the wind interaction in the CEE polar "channels" (the cored-apple structures in density). In the low momentum case, shocked fast wind accumulates in central regions of CEE ejecta. Inhomogeneities in the "walls" of the CEE ejecta, which are not symmetric about the equatorial plane, lead to the shocked fast wind being becoming trapped/diverted in local density pockets in the lower half of the domain compared to the upper half. Thus while the shocked fast wind is capable of pushing through the upper CEE channel to form an extended bipolar lobe, the fast wind in the lower channel is blocked for some time. By the time it escapes from the lower channel to form a jet, it significantly lags behind the upper lobe.

Note that the combination of the temperature and density maps show that once they form, the lobes are, just as in Model A, composed of high temperature, low density shocked fast wind material.

## 4. Conclusions

Using full CEE simulations as input we have performed AMR simulations that track the interaction of a fast wind from a central source (the compact binary) and a strongly aspherical CEE ejecta. We find that CEE ejecta provide the strong aspherical density environment needed in classic GISW models to produce strongly bipolar outflows such as those seen in some PPN and PN. Thus our models but confirm and extend the original emphasis on CEE in PN shaping.

We note that one significant new result is the ability of full 3-D CEE models to drive symmetry braking in GISW bipolar outflows. Our models not only show differences between top and bottom lobes in terms of inhomogenties, they also showed a difference in the size and extent of the upper and lower lobe. If this result continues to be obtained in more detailed and extensive modeling then it may offer a natural explanation for objects like OH231+8+04.2 and other highly asymmetric bipolar nebula.

Acknowledgments: This paper is supported the Extreme Science and Engineering Discovery Environment (XSEDE), supported by National Science Foundation grant number ACI-1548562. (through XSEDE allocation TG-AST120060) and with funding by Department of Energy grant GR523126, the National Science Foundation grant GR506177, the Space Telescope Science Institute grant GR528562 and NASA grants HST-15044 and HST-14563

Author Contributions: ZC and TR ran the simulations shown in this paper. AF,EB,JN and ODM helped in the set-up and analysis of the models

**Conflicts of Interest:** The authors declare no conflict of interest